# A GENERAL MODEL OF VEHICLE ROUTE GUIDANCE SYSTEMS BASED ON DISTRIBUTIVE LEARNING SCHEME

**Ke Wan, Zuo Zhang, Zhiquan Chen**
Ke Wan, Zuo Zhang and Zhiquan Chen are with the Department of Automation, Tsinghua University, Beijing 100084, P.R.China.
Email:wan-k03@mails.tsinghua.edu.cn

## ABSTRACT

Dynamic traffic assignment and vehicle route guidance have been important problems in ITS for some time. This paper proposes a new model for VRGS, which takes into consideration of the information propagation, user selection and information reaction. Parameter $p$ is then defined as the updating weight for computing cost of traffic based on a distributive learning scheme. $p$ is calculated through a function which denotes information propagation over time and space and the function needs further optimization. Comparison to static traffic assignment, DTA and feasible strategies are given, and future work is also stated.

## INTRODUCTION

Route guidance has raised much attention in ITS applications as one of the most important service types of traffic information system. As dynamic traffic assignment is the theory backbone of route guidance, its research has received attention in several decades. Wardrobe(1952)[1] proposed the first principle and second principle of traffic assignment and define the User Equilibrium(UE) state and System Optimum(SO) state for general network. Daganzo and Sheffi (1977)[2] gave the definition of Stochastic User Equilibrium. These led to enormous traffic assignment studies. However, many of them have the same underlying assumption "travelers have access to perfect information prior to setting out on a trip, they thus make a path choice decision at their origin and follow it to their destination"(**3**). It may actually handicap the dynamic performance of the assignment methods. Also, heavy computational load is another drawback of many DTA methods, which hampers the application of them.

On the other hand, due to the implementation of traveler information service, traffic network has become highly dynamic system and it brings new problems. For example, hunting phenomenon is well-known in route guidance applications and in simulation. With higher prevalence of in-vehicle information systems traffic jams will shift among several sites, for all drivers who get the same global information take the same action and there is lag in the information. The problem of the time lag can not be settled because of editing global information(**4**). A way of route decision by global traffic information and local information based on communication between drivers is proposed to avoid this(**5**). Jing Dong's simulation (2003) (**6**) on DTA concluded that local feedback was better than global feedback in terms of ATT, which needs further explanation.

This paper takes the distributive information processing to model the effect of traffic information on traffic assignment. It gives an analyzing Model for route guidance system, which decomposes the whole process into information propagation, user selection and information reaction. Then we focus on information propagation, and the distributive learning scheme of travel cost is formulated together with the information propagation optimization problem and its principles. Main strategies are then discussed with this model. Also several strategy design prototypes with primary consideration for optimization are showed. This paper is organized as follows. Section 2 gives the analyzing model. Section 3 formulates the distributive learning scheme of travel cost and the optimization problem. Section 4 applies the theory to discuss main information propagation strategies. In Section 5, future research issues are discussed.

**ROUTE GUIDANCE FUNCTIONAL MODEL**

Route guidance is a complex process with the interaction of human and traffic information system Functionally divided, route guidance process may contain the following three steps(**7**):

**Information propagation:**

Traffic information such as the latest travel time in a certain link propagates through the system and traveler or system client needs to get this information to update their perceived travel cost. This kind of data is the base for later route and a good propagation strategy is fundamental to optimize system performance(**8**)(**9**)(**10**).

The information propagation process is a highly dynamic and distributive process, during which a piece of information emerges from where its origin occurring site and propagate over time and space. Here two parameters are essential to define its affluence on the network:1). Influence at a certain time-space point. This denotes the power of the information to change a traveler's routing decision at a specified point.2). Total influence during propagation. This denotes the total effect of the information. In many studies information propagation strategy acts as the underlying assumption，they just give cursory assumptions and focus on information reaction to assign traffic. For example, many dynamic traffic assignment studies deal with network status change under the perfect information assumption mentioned in Section 1. However, it has showed some faults in Jing Dong's work(2003)(**6**).

Jing Dong gave simple comparison of local feedback and global feedback in the simulation result. With user selection and information reaction the same, the range of information propagation changes the system performance sharply, and the global feedback is worse than the local feedback, see Figure 1. There is no explanation for it in her work and we propose from this result that to assume the global perfect information propagation is natural but not optimal. The simple parameter guidance ratio or market penetration is never enough to address the mass behavior under wide prevalence of route guidance service.

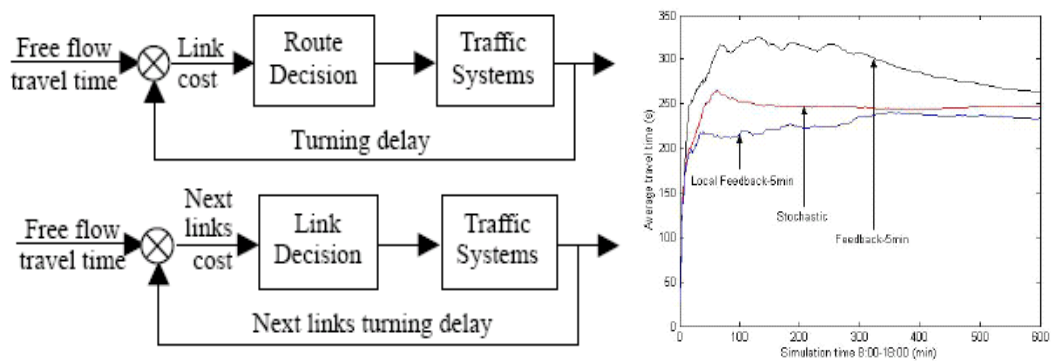

**Figure 1 Global feedback and Local feedback**

Moreover, from other study we can find analogous demand. Predictive feedback takes simultaneous simulation to predict future traffic state and generalize routing so as to approach dynamic user equilibrium. However, the problem lies that the further the destination from the traveler, the longer simulation time is in need. The simulation error and cost will increase sharply as the simulation time prolongs. And if we keep the simulation time constant for different destinations, then information propagation strategy design is necessary.

For the consideration above, we highlight information propagation as one of the main steps of routing guidance service and therefore it should be considered in dynamic traffic assignment problems. Information propagation strategy represents the dynamic nature of traffic under guidance and should be carefully studied and explained.

**User selection:**

User may choose whether to consider a piece of information or not when he makes the routing decision, or choose whether to follow the routing advice generalized by system client or not. This can be viewed as a multi-dimension-decided selection probability, which denotes the choice probability under certain contents of user, traffic status and service level.

$$P_{sel} = F(X_{serv}, X_{tra}, X_{user}) \tag{1}$$

$P_{sel}$ denotes the probability of selectivity under this context;

$X_{serv}$ denotes the characterizing vector of traffic information service;

$X_{tra}$ denotes the characterizing vector of the current traffic status;

$X_{user}$ denotes the characterizing vector of the traveler.

There are many field investigations on user behavior**(11)**. However, they mainly mix user selection with information reaction taken by users. Here we deal with them separately. User selection is the most interactive and changeful process in routing guidance. It can only be got through field investigation, and there is little chance for improvement. On the other hand, improvement of information reaction by users can enable the further optimization of system performance.

**Information reaction:**

This is the routing methods after getting traffic information. It can be done by travelers or just system client. Traditionally, it is always the focus of research and it may be achieved in the following ways: to follow mathematical traffic assignment**(12)(13)(14)(15)**, to follow advices of equilibrium indication feedback**(16)** and minimum travel cost routing**(17)**. Different information reaction strategies may be related to certain service types, require different information propagation strategies as assumptions or pre-steps, and have different goals (UE or SO), which may have conflicting results, see Table 1.

# A GENERAL MODEL OF VEHICLE ROUTE GUIDANCE SYSTEMS

Table 1 Comparison of different information reaction strategy

|  | Mathematical traffic assignment | Equilibrium indication feedback | Minimum travel cost routing |
|---|---|---|---|
| Information propagation | Perfect information acquisition | Underlying in Feedback parameter selection | Mainly Perfect information acquisition |
| User selection | Low | Low | **High** |
| Main goal | U.E S.O (Exact) | U.E (Approximate) | **Depend on information propagation strategy used** (Approximate) |
| Dynamic performance | Not satisfactory( partly due to the Information propagation assumption) | **Depend on information propagation strategy used** | **Depend on information propagation strategy used** |
| Main service type | Prescriptive | Prescriptive | Prescriptive Descriptive(adopted by user himself) |
| Calculation Load | Heavy | Small | Moderate |

Diagram 1 shows the information propagation design is essential to dynamic performance of routing system. And if a good information propagation strategy is designed, minimum travel cost routing, which has the highest user selection ratio will be best routing reaction methods, in spite of the type of service (descriptive/prescriptive).

User selection and information reaction may change order and be finished by either users or system client, under different routing service types ( descriptive / prescriptive). In descriptive routing, user selection and information reaction are both done by users, and user selection comes first. In prescriptive routing, information reaction is done by the system client and users select the route generalized later. The general procedures are shown in Figure2.

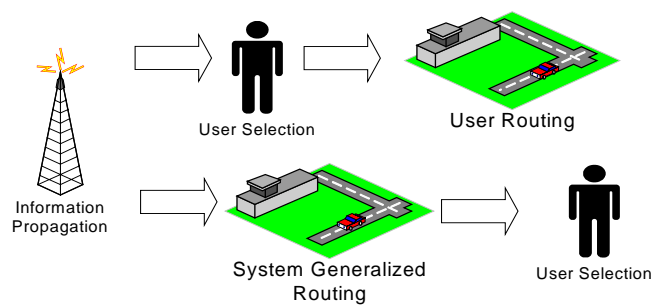

A GENERAL MODEL OF VEHICLE ROUTE GUIDANCE SYSTEMS

**Figure 2 Descriptive routing vs. Prescriptive routing**

## THE DISTRUBUTIVE LEARNING SCHEME AND OPTIMIZATION PROBLEM FORMULATION

In this section, we further focus on information propagation strategy design. We first model the generalized formulation of the user information upgrading process as follows, and the analysis of propagation strategies is then based on this model.

### The distributive learning scheme for traffic cost

The mass travelers learn the traffic cost at distributive sites. At each decision node, they use their perceived travel cost to generalize their routing. The perceived travel cost is calculated as follows:

$$C_{new}^{Per} = C_{old}^{Per} * (1-p) + C_{new} * p \tag{2}$$

$$p = F(X) = F(x, t, f, v, \ldots) \tag{3}$$

Define:

$C_{new}^{Per}$ denotes the old travel cost for a link perceived by a traveler;

$C_{old}^{Per}$ denotes the new travel cost for a link perceived by a traveler;

$C_{new}$ denotes the new travel cost for a link;

$p$ denotes the learning weight of the actual travel cost for a certain l

ink in the user perceived travel cost. It is the function of the
following variables;
$x$ denotes the radius of area for a piece of information about certain
travel cost change to propagate;
$t$ denotes the affecting duration for that travel cost;
$f$ denotes information upgrading frequency;
$v$ denotes the velocity of the propagation. It is mainly decided by the
system, and is somewhat related to $x$.

Here, the relation between the learning scheme and information propagation is as follows:

$p = F(X)$ denotes the relation between information propagation and the learning scheme. The learning scheme denotes the process of information updating from the side of users, which is direct related to information propagation from the side of system. How to calculate $p$ for a piece of traffic information to a certain user is the crucial issue of system design and needs optimization.

Besides, as a detailed formulation for the information propagation process, distributive learning scheme can represent the two important parameter of it:

$p = F(X)$ denotes the effect of information on routing for certain time-space point. The calculation function $F(X)$ is called information propagation function. Here we define the phase difference of two strategies:

For two strategies $S_1 : p_1 = F_1(X)$ and $S_2 : p_2 = F_2(X)$ If for a certain time-space point $(x,t)$, the relation: $p_1 = F_1(X) > p_2 = F_2(X)$ holds, then $S_1$ is called phase-lead over $S_2$, while phase-lag for $S_2$ over $S_1$ respectively.

The integration of $p = F(X)$ over time and space, which is

$$\iint_{x\,t} p\,dtdx = \iint_{x\,t} F(X)\,dtdx \tag{4}$$

denotes the total influence to the traffic network for a certain piece of information.

Furthermore, we have the equivalence measurement of two routing strategies based on this learning model: **[Equivalence law]**

Two different strategies designed which have the same integration value have cursory equivalent system performance. And the final performance difference is further determined by the phase difference between the two. That is:

For $S_1 : p_1 = F_1(X)$ and $S_2 : p_2 = F_2(X)$, if $\iint_{x\,t} F_1(X)dtdx = \iint_{x\,t} F_2(X)dtdx$ holds, then $\eta(S_1) \approx \eta(S_2)$. Furthermore, we have $\eta(S_1) - \eta(S_2) \longleftrightarrow \|F_1(X) - F_2(X)\|_{dis}$. Here:

$\eta(S)$ denotes the performance measure of a routing strategy $S$, such as ATT.

$\|F\|_{dis}$ denotes the measure of an information propagation function, and it can be defined by norm theory.

**The optimization problem of information propagation**

With the above model, the optimization problem is as follows:

# A GENERAL MODEL OF VEHICLE ROUTE GUIDANCE SYSTEMS

Problem Formulation: (**Information propagation function optimization**)

Keep user selection condition and information reaction strategy constant, and select appropriate $F(X)$ to minimize the certain performance measures (ATT and etc.) in equilibrium state and the time to get the equilibrium.

And the following principles are given to constrain and guide the function optimization:

Principle 1: (**Finite condition**)
The route guidance system can amplify this integration compared to that without the system, but its amplification is finite.

$$\int_x \int_t f(X)dtdx < \int_x \int_t F(X)dtdx < \infty \qquad (5)$$

Here $f(X)$ denotes the actual variation of the influence weight of a piece of information over time and space. The $F(X)$ got from field investigation is a good reference for this function.

If $F(X)=1$ it obviously violates this principle, for its integration tend to be infinite.

Principle 2: (**Phase inclination condition**)

$p = F(X)$ should represent the phase need change of traffic information over space and time.

$$\|F(X) - f(X)\|_{dis} \to 0 \qquad (6)$$

This principle indicates that the function should incline to the true variation of the influence weight of a piece of information over time and space. No distinct difference on phase is allowed for any time-space point. If $F(X)=(\Gamma(x+\Delta x) - \Gamma(x))$, it obviously violates this principle, for it never changes to time or space except the step point and phase difference is distinct. ($\Gamma(x)$ denotes the step function)

The two principles are necessary conditions to identify the solution to the optimization problem above. The first is a total affluence constrain and second is a strong equivalent condition. And the better an information propagation strategy satisfies these principles, the better its performance should be. This optimization problem is the way to relax the usual assumption on information propagation and helps to get a better system performance, both static and dynamic.

A GENERAL MODEL OF VEHICLE ROUTE GUIDANCE SYSTEMS

**STRATEGY ANALYSIS AND DESIGN**

We can then analyze traditional information propagation methods from this point of view, and can also design several possible strategies and evaluate their performance with the theory.

**Static assignment**, no information propagation: When no information propagation occurs, we have $p = 0$ in (2) and information never updates. The users base their routing on the static cost. No learning of new cost will induce to obvious bad routing result.

$$F(X) = 0 \qquad (7)$$

**DTA information propagation assumption**: The underling assumption of many DTA research is that "travelers have access to perfect information prior to setting out on a trip, they thus make a path choice decision at their origin and follow it to their destination". This equals to Natural Global Feedback proposed below. As what we will analyze, this underlying assumption hinders the final network performance sharply. However sophisticated mathematic methods they apply, they must first counteract the side effect of the assumption, which has been partly indicated in Section 2.

**Global feedback with a time gap**:

$$F(X) = 1*(\Gamma(t+\Delta t) - \Gamma(t)) \qquad (8)$$

$\Delta t$ is the duration for this information, its variation will change the impact of the information. Global feedback amplifies the actual influence of a piece of traffic information over $x$ to infinite, which finally violates Principle 1. Also the window function of $t$ can never incline to the true variation of the influence weight of a piece of information for a certain time, thus Principle 2 is violated. The feedback strategy mentioned in section 2 belongs to this category.

**Natural Global feedback strategy**:

$$F(X) = 1*F_t(t) \qquad (9a)$$

if we use negative exponential function to approximate the true attenuation function of traffic information influence over space, we get the following strategy: ($C_t$ is the expectation of the lasting time of a piece of information)

$$F(X) = 1*M_t^{-\frac{1}{C_t}t} \qquad (9b)$$

Just as Global feedback with a time gap, this strategy also amplifies the actual influence of a piece of traffic information over $x$ to infinite, which violates Principle 1. Here, the improvement lays that negative exponential function approximate the true variation of the influence weight of a piece of information for a certain time better. However, whether negative exponential function is the optimal needs to be justified, which will be discussed in

Section 5.

**Local feedback strategy with a time gap**:

$$F(X) = (\Gamma(0) - \Gamma(x)) * (\Gamma(t + \Delta t) - \Gamma(t)) \quad (10)$$

$x$ denotes the radius of the area a piece of information propagates. This strategy may satisfy Principle 1 and be optimized through parameter adjusting but it will always violate Principle 2, because the window function of $x$ and $t$ can never incline to the true variation of the influence weight of a piece of information for a certain space-time point. The local feedback strategy mentioned in section 2 belongs to this category. As the improvement that it obeys Principle 1, it has got better performance than the feedback strategy.

**Natural Local feedback strategy**:

$$F(X) = (\Gamma(0) - \Gamma(x)) * F_t(t) \quad (11a)$$

if we use negative exponential function to approximate the true attenuation function of traffic information influence over time, we get the following strategy:

$$F(X) = (\Gamma(0) - \Gamma(x)) * M_t^{-\frac{1}{C_t}t} \quad (11b)$$

This strategy may satisfy Principle 1 through parameter adjusting but it will always violate Principle 2 because the window function of $x$ can never incline to the true variation of the influence weight of a piece of information for a certain space point.

**Natural space-time approximating strategy**:

$$F(X) = F_x(x) * F_t(t) \quad (12a)$$

if we use negative exponential function to approximate the true attenuation function of traffic information influence over both time and space, we get the following strategy: ($C_x$ denotes the expectation of the radius of the area a piece of information propagates)

$$F(X) = M_x^{-\frac{1}{C_x}x} * M_t^{-\frac{1}{C_t}t} \quad (12b)$$

This strategy may satisfy both Principle 1 and Principle 2 to a better extent, without the limitation of window function. Obviously, it needs further optimization.

**CONCLUSION AND FUTURE WORK**

Distributive learning of traffic cost is effective to applications. It is instinctively dynamic compared to centralized methods and the processing is further optimized beyond simple global or local feedback. The strategies mentioned in this paper are only main types of functions based on primary optimization consideration, without exact parameter optimization which may be different for concrete network status.

Several methods are feasible under this concept to further theory study and sophisticated applications. 1)Theory optimization: Add the optimization problem into traditional DTA problems, and solve the following sub problems together: First, functional optimization for the best information propagation function, including independent variable set identification, function type selection and parameter optimization; Second, information reaction strategy design for a variant OD network.2) Simulation experiment: Hold the basic traffic behavior, user selection and information reaction strategies constant, use simulation to show the performance change due to the change of information propagation and get further optimization(**18**). However, simulation optimization generally can not optimize the function type but the function parameters, so that the optimization across different function types needs continuing trial. 3) Field investigation: this can be done to acquire the influence power data from field test stand for general human reaction, and is a good reference for system design of $F(X)$. However, it can not necessarily be the optimal.

Appropriate information propagation strategy is fundamental to good performance of route guidance system. Although certain simple assumptions on it are convenient for system modeling, they bring side effects simultaneously. Only by formulating it as an optimization problem and solving it, will it be possible to raise the performance of routing guidance system further, and to connect traffic theory study with concrete applications better.